\documentclass{PoS}
\usepackage{graphicx}

\newcommand{\be}{\begin{equation}}
\newcommand{\ee}{\end{equation}}
\newcommand{\bea}{\begin{eqnarray*}}
\newcommand{\eda}{\end{eqnarray*}}
\newcommand{\bda}{\begin{eqnarray}}
\newcommand{\eea}{\end{eqnarray}}
\newcommand{\FC}{\;,}
\newcommand{\FD}{\;.}
\newcommand{\srm}[1]{\textrm{\scriptsize #1}}
\newcommand{\MSbar}{$\overline{\textrm{MS}}$}

\newcommand{\VEC}[1]{\ensuremath{\mathbf{#1}}}
\newcommand{\eq}[1]{(\ref{#1})}

\PoS{PoS(LAT2005)126}

\title{Pion form factor with chirally improved fermions \footnote{For the 
Bern-Graz-Regensburg (BGR) collaboration.}}

\ShortTitle{Pion form factor with chirally improved fermions }

\author{\speaker{Stefano Capitani}\thanks{
        Supported by Fonds zur F\"orderung der Wissenschaftlichen Forschung 
        in \"Osterreich, Project P16310-N08.}\\
        Karl-Franzens-Universit\"at, Graz, Austria\\
        E-mail: \email{stefano.capitani@uni-graz.at}}

\author{Christof Gattringer\\
        Karl-Franzens-Universit\"at, Graz, Austria\\
        E-mail: \email{christof.gattringer@uni-graz.at}}

\author{C. B. Lang\\
        Karl-Franzens-Universit\"at, Graz, Austria\\
        E-mail: \email{christian.lang@uni-graz.at}}

\abstract{
We present results for Monte Carlo calculations of the electromagnetic vector
and scalar form factors of the pion in a quenched simulation. We work at a
lattice spacing of 0.15 fm and use two lattice volumes up to a spatial size
of 2.4 fm. The pion form factors in the space-like region are determined
for pion masses down to 340 MeV.}

\FullConference{XXIIIrd International Symposium on Lattice Field Theory\\
		 25-30 July 2005\\
		 Trinity College, Dublin, Ireland}

\begin{document}

\section{Introduction}

We present a first-principles calculation of off-forward matrix elements of
operators that measure the vector and scalar form factors of the pion,
utilizing chirally improved fermions \cite{Ga01,GaHiLa00}. 
The vector form factor $F_\pi$ is defined by
\be\label{eq1}
\langle \pi^+ (\VEC{p}_f) |\, V_\mu \,| \pi^+ (\VEC{p}_i) \rangle_\srm{cont}
= (p_f+p_i)_\mu \, F_\pi (Q^2)\FC
\ee
where $Q^2 = (p_f-p_i)^2$ is the space-like invariant momentum transfer 
squared, and $V_\mu = 2/3 \; \overline{u} \,\gamma_\mu \,u -
                   1/3 \; \overline{d} \,\gamma_\mu \,d$ is the vector current.
Electromagnetic gauge invariance implies the constraint $F_\pi (0) = 1$.
The mean square charge radius is defined as
\be
\langle r^2 \rangle_\srm{v} = - 6 \, \frac{d}{dQ^2} \, F_\pi (Q^2) 
\Big|_{Q^2=0}\FD
\ee
The scalar form factor $\Gamma_\pi$ is given by the matrix element of the 
scalar operator
\be
\langle \pi^+ (\VEC{p}_f) | \,m_u\,\overline{u}\, u + m_d\,\overline{d} 
\,d\,| \pi^+ (\VEC{p}_i) \rangle = \Gamma_\pi (Q^2)\FD
\ee
From chiral perturbation theory one knows that the scalar form factor
at $Q^2=0$ is the so-called sigma term which behaves like
$\Gamma_\pi (0) \sim  M_\pi^2$ near zero momentum transfer. 
The scalar radius $\langle r^2 \rangle_s$ can be obtained from
\be\label{defscalarradius}
\frac{\Gamma_\pi (Q^2)}{\Gamma_\pi (0)} =
1 - \frac{1}{6} \, \langle r^2 \rangle_s \, Q^2 + {\cal O}\big(Q^4\big)\FD
\ee
For a detailed discussion of chiral perturbation theory in this context see 
\cite{AnCaCo04}.

\section{Strategy}

In order to compute the form factors, we need to evaluate off-forward matrix 
elements (at several transferred momenta), i.e., the expectation values of
\be
{\mathrm{Tr}} \, \left(
\sum_{\VEC{y};y_0=\tau} \, e^{i\, \VEC{q}\cdot\VEC{y}} \, S(0,y) \; O \;
\sum_{\VEC{x};x_0=t}   \, e^{-i\, \VEC{p}\cdot\VEC{x}} \, S(y,x) \, \gamma_5 
\, S(x,0) 
\, \gamma_5  \right)\FC
\ee
where $S(y,x)$ is the quark propagator from $x$ to $y$ and $O$ denotes
the operator inserted at $y$. We use the notation $\VEC{p}\equiv\VEC{p}_f$, 
$\VEC{r}\equiv\VEC{p}_i$ and momentum transfer $\VEC{q}=\VEC{p}-\VEC{r}$.
These matrix elements have been evaluated by using the sequential source 
method where the sequential propagator can be easily computed by an additional 
inversion of the Dirac operator for each choice of the final momentum.
 
We extract physical matrix elements by computing ratios
of 3-point and 2-point correlators:
\be
R(t,\tau;\VEC{p},\VEC{q})  = 
\frac{\langle P(t;\VEC{p})\,O(\tau;\VEC{q})\,\overline{P}(0;\VEC{r}) \rangle
}{\langle P(t;\VEC{p})\,\overline{P}(0;\VEC{p}) \rangle} \,
\sqrt{ \, \frac{\langle P(t;\VEC{p})\,\overline{P}(0;\VEC{p})\rangle \,
                \langle P(\tau;\VEC{p})\,\overline{P}(0;\VEC{p})\rangle \,
                \langle P(t-\tau;\VEC{r})\,\overline{P}(0;\VEC{r})\rangle }{
                \langle P(t;\VEC{r})\,\overline{P}(0;\VEC{r})\rangle \,
                \langle P(\tau;\VEC{r})\,\overline{P}(0;\VEC{r})\rangle \,
                \langle P(t-\tau;\VEC{p})\,\overline{P}(0;\VEC{p})\rangle }}\FC
\label{eqcorr}
\ee
where $P$ denotes the pseudoscalar interpolator.
We keep the sink fixed at time $t$ and change the timeslice $\tau$ where 
the operator $O$ sits (scanning a range of timeslices).
$R(t,\tau;\VEC{p},\VEC{q})$ should exhibit two plateaus in $\tau$, 
for $0 \ll \tau \ll t$ and $t \ll \tau \ll T$.  
Note that the terms under the square root become trivial for $\VEC{q}=(0,0,0)$.

\begin{table}[t]
\begin{center}
\begin{tabular}{c|cccc}
\hline
\hline
$16^3\times 32$ & 0.74~GeV & 1.04~GeV & 1.28~GeV & 1.48~GeV \\
$12^3\times 24$ & 0.98~GeV & 1.39~GeV & 1.71~GeV & 1.97~GeV \\
\hline
\end{tabular}
\end{center}
\caption{Values of the nonzero momentum transfers.}
\label{tab1}
\end{table}

We choose for initial and final momenta $|\VEC{p}_f | = | \VEC{p}_i | $
such that $E_{f} = E_{i}$ and the transferred 4-momentum is given by 
$Q^2 = |\VEC{q}|^2$. In this way one achieves for the electromagnetic 
form factor \cite{HeKoLa04} the cancellation of the kinematical factors 
in \eq{eq1} and
\be
\langle \pi (\VEC{p}_f) | V_\mu | \pi (\VEC{p}_i) \rangle_\srm{latt}
= \frac{1}{2\,\sqrt{E_f E_i}}
\langle \pi (\VEC{p}_f) | V_\mu | \pi (\VEC{p}_i) \rangle_\srm{cont}\FD
\ee
Indeed, with this choice of momenta, and using the $\mu =4$ component of 
the vector current, the overall factor becomes 
\be
\frac{E_f + E_i}{2\sqrt{E_f E_i}} = 1 .
\label{eqsimplification}
\ee
This cancellation is unfortunately no longer possible in the calculation 
of the scalar form factor, where a remaining multiplication by the quantity
$2\,\sqrt{E_f \,E_i}$ is still needed. 

We have made the choice $|\VEC{p}_f| = |\VEC{p}_i| = \sqrt{2}\, p_0$, where 
$p_0=2\,\pi/ (a\,L)$ denotes the smallest (spatial) momentum unit. Fixing the 
final (sink) momentum to $\VEC{p}_f=(1,1,0)\,p_0$, we get twelve values for 
the initial momentum. They give rise to four nonzero (and equidistant) values 
for the momentum transfer, $Q^2 = 2\,n \,p_0^2$ for $n=0,\,1,\,2,\,3,\,4$ 
(see Table~\ref{tab1}). We could also consider larger values of the module 
of the initial momentum. In this case, however, the statistical errors 
become much larger, and moreover we could not use Eq.~\eq{eqsimplification} 
anymore. A change of the momentum at the sink would require the expensive 
computation of a new set of sequential propagators.

In the Monte Carlo simulations additional numerical noise due to the
momentum projection of the sink and of the operator onto nonzero momentum
transfer $\VEC{q}$ arises. The 2-point correlators may even become negative
(within the statistical errors) on timeslices near the symmetry point. 
A reasonable choice was to put the sink at a timeslice smaller than $T/2$, 
such that the 2-point functions have smaller errors. In particular, we have 
put the sink at timeslice $t=8$, while the source sits at timeslice $t=1$.

\section{Simulation parameters}

We use the chirally improved Dirac operator \cite{Ga01,GaHiLa00,GaGoHa03a}, 
which constitutes an approximate Ginsparg-Wilson operator. The gauge action 
is the L\"uscher-Weisz tadpole improved action at $\beta=7.9$, corresponding 
to a lattice spacing of 0.148 fm (determined from the Sommer parameter). 
Although we expect improved scaling properties, here we do not study scaling 
behavior. The hadron interpolators are constructed from Jacobi-smeared quark 
sources and sinks.

We have done simulations using two volumes, $12^3\times 24$ (200 configurations
for each mass) and $16^3\times 32$ (100 configurations for each mass). 
This corresponds to spatial lattice sizes of $a\,L \sim 1.8\;$fm and $a\,L 
\sim 2.4\;$fm respectively. 

We work at several values of the bare quark mass; the corresponding meson 
masses are given in Table~\ref{tab2}. The jackknife method is used for 
estimating the statistical errors of our results.

\begin{table}[t]
\begin{center}
\begin{tabular}{c|ccc|cc}
\hline
\hline
&  $16^3\times 32$  &  $16^3\times 32$  & $16^3\times 32$ 
&  $12^3\times 24$  &  $12^3\times 24$  \\
\hline
$m$      &  0.02  $a^{-1}$   &  0.04  $a^{-1}$   &  0.06  $a^{-1}$   
         &  0.04  $a^{-1}$   &  0.06  $a^{-1}$   \\ 
$m_\pi$  &  342~MeV  &  471~MeV  &  571~MeV  &  474~MeV  &  575~MeV  \\ 
$m_\rho$ &  845~MeV  &  895~MeV  &  941~MeV  &  952~MeV  &  976~MeV  \\
\hline
\end{tabular}
\end{center}
\caption{Pion and rho masses at the various quark masses simulated
\cite{GaGoHa03a}.}
\label{tab2}
\end{table}

\section{Results}

\begin{figure}[b]
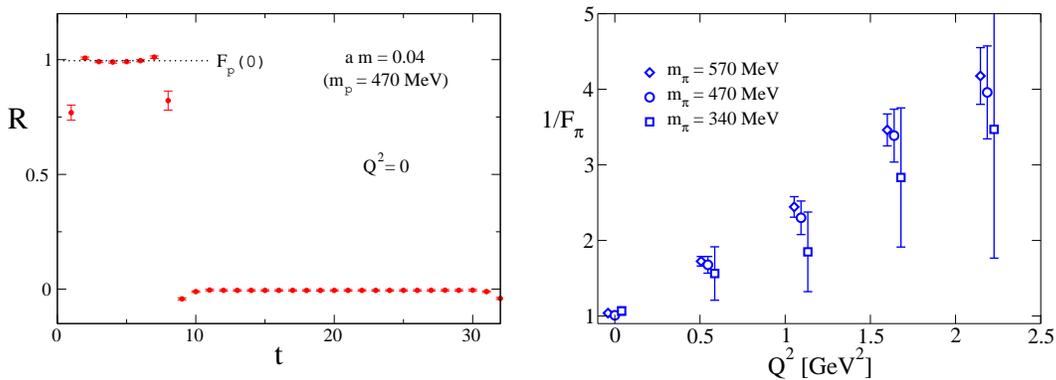

\begin{center}
\includegraphics*[height=5cm]{fig1a.eps} \quad
\includegraphics*[height=5cm]{fig1b.eps} 
\end{center}
\caption{(a) L.h.s.: The ratio $R(t,\tau;\VEC{p},\VEC{q})$ leading to 
the electromagnetic vector form factor at zero momentum transfer, for the 
$16^3\times 32$ lattice at $m_\pi = 470$ MeV.
(b) R.h.s.: The inverse vector form factor for the three available masses 
on the $16^3\times 32$ lattice. For presentation purposes the data points 
for different masses are slightly shifted horizontally.}
\label{fig1}
\end{figure}

In Fig.~\ref{fig1}a we show an example for the observed plateau behavior
of the ratio \eq{eqcorr}. Fitting the central points of the plateau then 
provides values for the electromagnetic form factor of the pion. 

When comparing the results to physical, renormalized quantities we have 
to multiply with renormalization factors relating the raw results to the 
\MSbar-scheme. For the chirally improved Dirac operator these have been 
determined in \cite{GaGoHu04} and all results we show are already converted 
to the \MSbar-scheme. Although the vector current is pointlike and not 
conserved, the value $Z_V=0.9586(2)$ turns out to be close to unity. 
The resulting $F_\pi (0)$ is not constrained to unit value, but comes very 
close as may be seen, e.g., in Fig.~\ref{fig1}a. This fact is also obvious 
from Fig.~\ref{fig1}b, where we plot the inverse of the vector form factor
determined for all the momentum transfer values and quark masses studied 
on the $16^3\times 32$ lattice.

The electromagnetic pion form factor in the time-like region is dominated 
by the $\rho$-meson. In the space-like region it may therefore be well 
approximated by a monopole form, $F_\pi (Q^2) = {m_V^2}/({m_V^2+Q^2})$, 
such that $1/F_\pi$ has the linear behavior indeed observed in 
Fig.~\ref{fig1}b. In that approximation the mean square radius is $6/m_V^2$.  

As the vector meson dominance (VMD) model is just an approximation one expects 
corrections due to other resonances and more-particle channels in the 
time-like region. These may be taken into account by additional pole terms. 
The data (cf.\ Fig.~\ref{fig1}b) does not really require such a multi-parameter
fit and for the presentation here we discuss only the results of a linear fit.

\begin{figure}[tp]
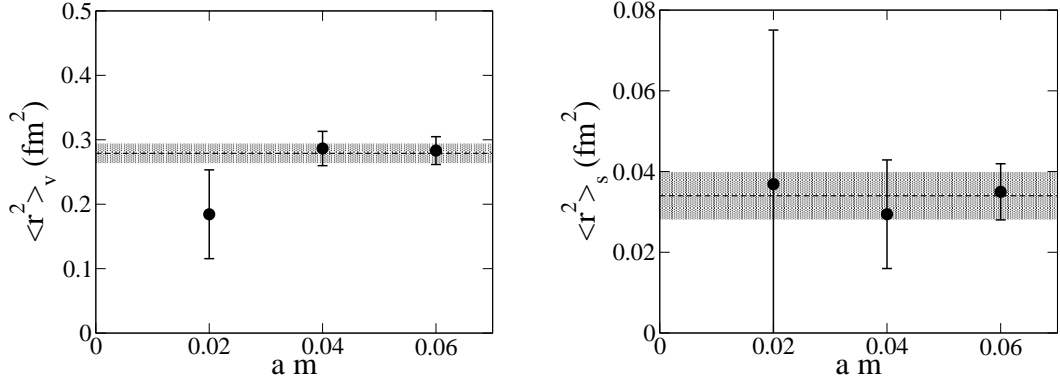

\begin{center}
\includegraphics*[height=5cm]{fig2a.eps} \quad \quad
\includegraphics*[height=5cm]{fig2b.eps} 
\end{center}
\caption{The mean square radius for the three available masses 
on the $16^3\times 32$ lattice for (a) the vector (l.h.s.) and 
(b) the scalar (r.h.s.) form factor.}
\label{fig2}
\end{figure}

The derivative of the form factor at $Q^2=0$ gives the charge radius, as shown 
in Fig.~\ref{fig2}a. Whereas the large mass results are compatible 
with a constant, the lowest mass is smaller but with a very large statistical 
error. The values are quite compatible with values from \cite{HeKoLa04} at 
comparable quark mass values. A fit to a constant gives 
$\langle r^2\rangle_\srm{v}=0.279(16)\;\textrm{fm}^2$. 
The current PDG average is $0.45(1)$ \cite{PDBook}.

The value of $\langle r^2\rangle_\srm{v}$ is in the VMD model inversely 
proportional to the mass squared of the $\rho$-meson. We obtain 
$m_\srm{VDM}^2=0.83(5)\;\textrm{GeV}^2$. 
This is close to the range of values obtained for $m_\rho^2$ in the BGR 
analysis \cite{GaGoHa03a} in the direct $\rho$-channel 
(e.g., $0.80\;\textrm{GeV}^2$ at $a\,m=0.04$). These large values for the 
$m_\rho$ mass might at least in part explain (via the VMD model) our low 
result for $\langle r^2\rangle_\srm{v}$.

The scalar form factor also has disconnected contributions which we disregard 
here (like it is done usually due to the related technical complications). 
We determine the scalar radius squared from a linear fit to the 
$Q^2$-dependence of the scalar form factor (now explicitely
normalized according to Eq.~\ref{defscalarradius}). The resulting values, 
shown in Fig.~\ref{fig2}b, are almost an order of magnitude smaller 
than for the electromagnetic case. We obtain an average for the scalar radius 
of $\langle r^2\rangle_s=0.034(6)\;\textrm{fm}^2$. It is also smaller than the 
values expected for full QCD.

In chiral perturbation theory the scalar radius is related to the 
quark mass dependence of the pion decay constant  
\be
f_\pi / f = 1 + \frac{1}{6} \, \langle r^2 \rangle_s \, M_\pi^2  +
\frac{13\, M_\pi^2}{192 \pi^2 f_\pi^2} + {\cal O}\big(M_\pi^4)\FC
\ee
and in Ref.~\cite{AnCaCo04} values around $\langle r^2 \rangle_s \sim
0.55-0.75\;\textrm{fm}^2$ were expected. Results for the pion decay constant 
in a recent full QCD lattice calculation \cite{AuBeDe04} lead to the value 
$\langle r^2 \rangle_s = 0.5 \pm 0.1~\textrm{fm}^2$. A corresponding analysis 
of the quenched BGR data \cite{GaHuLa05} gives again a small value  
$0.08-0.13\;\textrm{fm}^2$. This seems to imply that $\langle r^2 \rangle_s$
is very sensitive to quenching. 

\section{Conclusions}

Chirally improved fermions provide a valid framework for a first principles 
lattice study of the hadron structure. They are Ginsparg-Wilson-like fermions 
and less expensive than other types of chiral fermions. We have begun the 
study of hadron structure by investigating the form factors of the pion. 
The vector form factor is consistent with expectations in its general form, 
although the charge radius is 40\% smaller than in experiment. For the scalar 
form factor the scalar radius is much lower than one would expect from 
unquenched calculations. Both effects may be signals of quenching.

Further progress could be achieved - as usual - by using better statistics, 
more momenta, and larger lattices, both in lattice units (to study finite 
volume effects) and in physical units (to access lower transferred momenta).

Most computations were done on the Hitachi SR8000-F1 at the Leibniz 
Rechenzentrum in Munich, and we thank the staff for support.


\end{document}